%% file: main.tex
\documentclass[11pt]{article}

\usepackage[cm]{fullpage}
\usepackage[small,compact]{titlesec}
\usepackage[small,it]{caption}

\usepackage{amsfonts}
\usepackage{amsmath}
\usepackage{amssymb}

\usepackage{graphicx}
\usepackage{times}
\usepackage{url}

\newcommand{\sq}{\hbox{\rlap{$\sqcap$}$\sqcup$}}
\newcommand{\qed}{\hspace*{\fill}\sq}

\newcommand{\figScale}{0.55}

\begin{document}

\begin{titlepage}

\title{Ameba-inspired Self-organizing Particle Systems\\
{\Large (Extended Abstract)}}

\author{
	Shlomi Dolev$^1$, Robert Gmyr$^2$, Andr\'ea W.\ Richa$^3$, Christian Scheideler$^2$\\
	\\
	$^1$ Department of Computer Science,\\
	Ben-Gurion University of the Negev, Israel,\\
	dolev@cs.bgu.ac.il\\
	\\
	$^2$ Department of Computer Science,\\
	University of Paderborn, Germany,\\
	gmyr@mail.upb.de, scheideler@mail.upb.de\\
	\\
	$^3$ Department of Computer Science and Engineering,\\
	Arizona State University, USA,\\
	aricha@asu.edu
}

\date{}

\maketitle \thispagestyle{empty}

\begin{abstract}
Particle systems are physical systems of simple computational particles
that can bond to neighboring particles and use these bonds to move from one spot to another (non-occupied) spot.
These particle systems are supposed to be able to self-organize
in order to adapt to a desired shape without any central control.
Self-organizing particle systems have many interesting applications
like coating objects for monitoring and repair purposes
and the formation of nano-scale devices for surgery and molecular-scale electronic structures.
While there has been quite a lot of systems work in this area,
especially in the context of modular self-reconfigurable robotic systems,
only very little theoretical work has been done in this area so far.
We attempt to bridge this gap by proposing a model inspired by the behavior of ameba that allows
rigorous algorithmic research on self-organizing particle systems.
\end{abstract}

\bigskip

\centerline{{\bf Keywords}: Particle systems; self-organization; nano-structures; nano-computing.}

\end{titlepage}

\section{Introduction}
\label{sec:introduction}
Over the next few decades, two emerging technologies---microfabrication and cellular engineering---will
make it possible to assemble systems incorporating myriads of simple information processing units at almost no cost. Microelectronic mechanical components have become so inexpensive to manufacture
that one can anticipate integrating logic circuits, microsensors, actuators, and communication devices
on the same chip to produce particles that could be mixed with bulk materials,
such as paints, gels, and concrete \cite{AAC+00}.
Imagine coating bridges and buildings with smart paint
that senses and reports on traffic and wind loads and monitors structural integrity.
A smart-paint coating on a wall could sense vibrations, monitor the premises for intruders, and cancel noise.
There has also been amazing progress in understanding the biochemical mechanisms in individual cells
such as the mechanisms behind cell signaling \cite{cellSignaling} and cell movement \cite{AE07}.
Recent results have also demonstrated that, in principle,
biological cells can be turned into finite automata \cite{BPEA+01} or even pushdown automata \cite{KSP12},
so one can imagine that some day one can tailor-make biological cells to function as sensors and actuators,
as programmable delivery devices, and as chemical factories for the assembly of nano-scale structures.
Particularly interesting applications for this technology would be the construction of
molecular-scale electronic structures and of nano-scale devices for surgery as well as cancer treatment \cite{WBX11}.

Yet fabrication is only part of the story.
One can envision producing vast quantities of individual computing elements---whether
microfabricated particles or engineered cells---but
research on how to use them effectively is still in its infancy.
The opportunity to exploit these new technologies poses a broad conceptual challenge
that was coined by Abelson, Knight, Sussman, et al.\ as \textit{amorphous computing} \cite{AAC+00}.
In amorphous computing one usually assumes that there is a very large number of locally interacting computing elements,
called \textit{computational particles}, that may form an arbitrary initial structure.
The particles are possibly faulty, sensitive to the environment, and may produce various types of actions
that range from changing their internal state to communicating with other particles,
moving to a different location, or even replicating (to mimic biological cell replication).
Each particle has modest computational power and a modest amount of memory.
The particles are not synchronized, although it is usually assumed that they compute at similar speeds
(as long as they are non-faulty), since they are all fabricated by the same process.
The particles are all programmed identically,
but they usually have means for storing local state and for generating random numbers,
which allows them to differentiate over time.
In general, the particles do not have any a priori knowledge of their positions or orientations.

We propose a new ameba-inspired model for computational particles representing finite automata
that form a connected structure with the help of local bonds.
The particles cannot move through other particles or in open space,
and while they move they cannot drag other particles with them due to limitations on their energy and strength.
So if they wish to form a particular pattern,
then the only way to achieve that is through individual movements of particles along the surface of a particle structure
(which could be done by releasing and engaging bonds to static particles
in a similar way biological cells are moving \cite{AE07})
until they have reached the desired location.
This is continued until the desired pattern has been reached.
Although it is clear that this can have many interesting applications like the repair of structural damages,
theoretical research on our type of particle systems is basically non-existent,
so there is a dire need to provide a solid theoretical base.

\section{Related Work}
\label{sec:relatedWork}
Self-organizing systems have been studied in many contexts.
One can distinguish here between active and passive systems.
In active systems, there are computational particles that can control the way they act and move,
while in \text{passive systems} the particles either do not have any intelligence at all,
or they may have limited computational capabilities but they cannot control their movements.
Examples of algorithmic research on passive systems are
DNA computing \cite{Adl94,BDLS96,CDBG11,DPSS11,NKC03,WLWS98},
population protocols \cite{AAD+06}, and slime molds \cite{BMV12,LTT+10,WTTN11}.
We will not describe these models in detail as they are only of little relevance for our approach.
Examples of active systems are self-organizing networks as well as swarm robotics and modular robotic systems.

\textit{Self-organizing networks} have been studied in many different contexts.
Networks that evolve out of local, self-organizing behavior have been heavily studied in the context of
\textit{complex networks} \cite{complexNetworks} such as small-world networks \cite{BA99,Kle00,WS98}.
However, whereas a common approach for the complex networks field is to study
the global effect of given local interaction rules,
we are interested in developing local interaction rules in order to obtain a desired global effect.
Also so-called \textit{self-stabilizing overlay networks} have been studied,
see, for example, \cite{Dol00} for an in-depth treatment.
While the proof techniques used in this area promise to be useful also for self-organizing particle systems,
the constraints on particle systems are much more severe than on nodes in overlay networks,
which can freely change their interconnections and which have no limitations on their computational power.

In \textit{swarm robotics} it is usually assumed that there is a collection of autonomous robots
that have a limited sensing and communication range and that can freely move in a given area.
Surveys of recent results in this area can be found, e.g., in \cite{Ker12,McL08}.
While the analytical techniques developed in this area are of some relevance for this work,
the underlying model differs significantly as we do not allow free movement of particles.
While swarm robotics focuses on inter-robotic aspects in order to perform certain tasks,
the field of \textit{modular self-reconfigurable robotic systems} focuses on intra-robotic aspects
such as the design, fabrication, motion planning, and control of autonomous kinematic machines with variable morphology.
Since the field started with the development of CEBOT by Toshio Fukuda \cite{FNKB88},
a growing number of research groups have become actively involved in modular robotics research \cite{FR03,YSS+07}.
\textit{Metamorphic robots}  form a subclass of self-reconfigurable robots that shares the characteristics of our model
that all particles are identical and that they fill space without gaps \cite{Chi94}.
The hardware development in the field of self-reconfigurable robotics has been complemented
by a number of algorithmic advances (e.g., \cite{BKRT04,CP96,Ost04,WWA04}),
but so far mechanisms that scale to hundreds or thousands of individual units are still under investigation,
and no rigorous theoretical foundation is available yet.
So although the advances in this area are relevant for the feasibility of our model,
the area does not provide algorithmic and analytical techniques that are directly applicable
to our self-organizing particle systems.

\section{The Artificial Ameba}
\label{sec:theArtificialAmeba}
In the following, we present our ameba-inspired model for particle systems in two dimensions.
In our model, particles are of hexagonal shape such that they can fill space without gaps.
The number of particles in the system is not known by any individual particle and particles are anonymous.
Furthermore, particles do not know their position or orientation (no-compass model, e.g., \cite{EP07}).
However, the particles are assumed to be able to distinguish clock-wise from counter-clock-wise.
Particles form bonds with their immediate neighbors and the particles in a particle system
have to form a connected structure at all times.
Computationally, particles act like probabilistic finite automata.
Note, that this decision does not restrict the number of particles in the system as particles have no global knowledge.
The particles act in synchronous rounds.
A particle uses the configurations of neighboring particles together with its own configuration
to compute its next action.
Locomotion of particles is realized by sequences of expansion and contraction.
Both, the ability to form bonds and the locomotion of particles,
are inspired by a behavior called \textit{cell crawling}
that can be found in ameba \cite{Fuk02} and cells like macrophages \cite{AE07}.
Our decision for hexagonal particles is a consequence of the way particles connect and move
since other regular polygons that tile two-dimensional space, namely the triangle and the square,
introduce certain technical problems in this context.
As an example, square particles might require diagonal connections for one particle to move around another particle;
however, diagonal connections might be hard to implement since diagonal particles meet only in one point.
Finally, particles are allowed to replicate and dissolve
which is inspired by cell division and cell death.

\begin{figure}
	\begin{minipage}{0.45\textwidth}
		\centering
		\includegraphics[scale = \figScale]{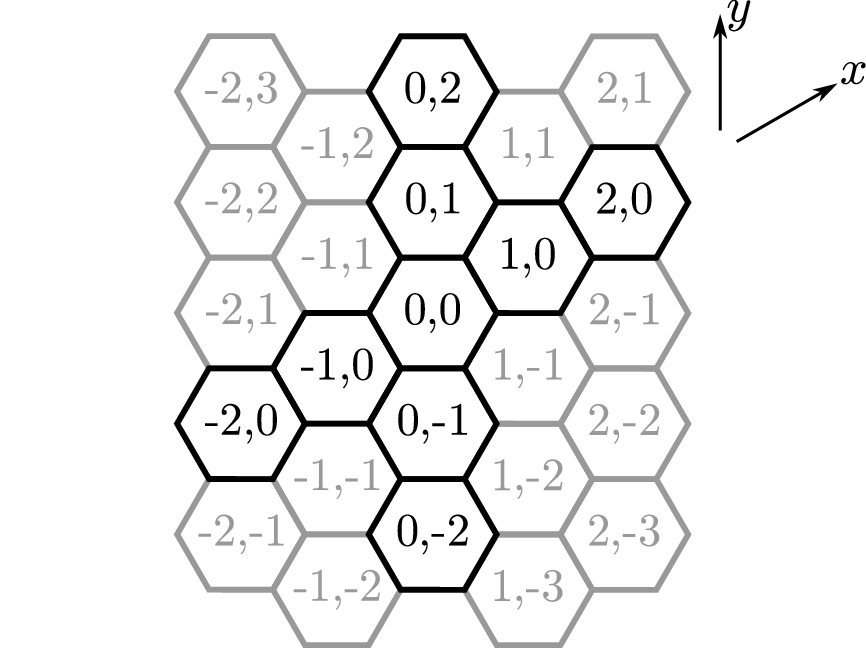}
		\caption{
			Coordinates in the hexagonal grid.
			The cells that constitute the coordinate axes are shown in black.
		}
		\label{fig:grid}
	\end{minipage}%
	\hspace{0.1\textwidth}%
	\begin{minipage}{0.45\textwidth}
		\centering
		\includegraphics[scale = \figScale]{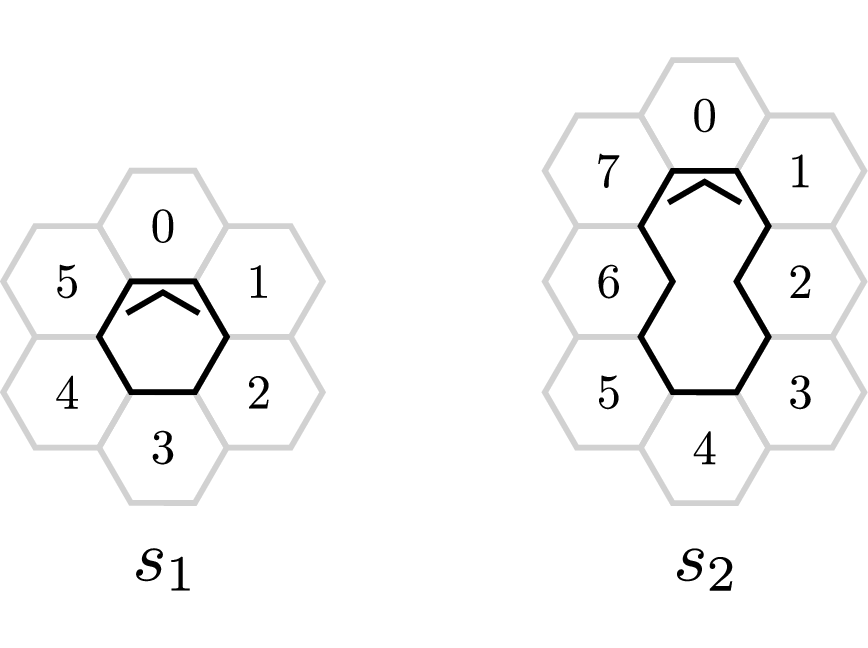}
		\caption{
			The two shapes of particles.
			The neighboring cells (gray) are numbered according to the orientation of the particle (arrow);
			the cell labeled with number $i$ is the neighbor $n_i(p)$ of the respective particle $p$.
		}
	\label{fig:shapes}
	\end{minipage}	
\end{figure}

Formally, a particle $p$ is a tuple $(Q, \delta, q_0)$ where $Q$ is a set of states,
$\delta$ is a probabilistic transition function, and $q_0 \in Q$ is the \textit{start state}.
A particle system $P$ is an ordered set of particles.
The particles are positioned on an infinite hexagonal grid.
We assign coordinates from $\mathbb{Z}^2$ to this grid as depicted in Figure~\ref{fig:grid}.
This coordinate system is adopted from \cite{WWA00}
and is a modification of the coordinate system presented in \cite{Chi94}.
An element of the grid is called a \textit{cell} $c \in \mathbb{Z}^2$.
A \textit{direction} is an element $d \in \mathbb{Z}_6$ where $\mathbb{Z}_6$ are the integers modulo $6$.
We define the \textit{neighbor} $n_d(c)$ of a cell $c$ in direction $d$ as follows.
\[
n_d(c) = 
\begin{cases}
	\begin{array}{llcccllcccll}
	c + ( 0,  1) & \text{, if } d = 0 &&& c + ( 1,  0) & \text{, if } d = 1 &&& c + ( 1, -1) & \text{, if } d = 2\\
	c + ( 0, -1) & \text{, if } d = 3 &&& c + (-1,  0) & \text{, if } d = 4 &&& c + (-1,  1) & \text{, if } d = 5
	\end{array}
\end{cases}
\]
Accordingly, the \textit{neighborhood} of a cell $c$ is defined as $N(c) = \{n_d(c) \mid d \in \mathbb{Z}_6\}$.
Every particle assumes a \textit{shape} $s \in S = \{s_1, s_2\}$, see Figure~\ref{fig:shapes}.
We denote the set of cells occupied by a particle $p$ as $C(p)$; accordingly we have $|C(p)| \in \{1, 2\}$.
We call a cell \textit{occupied} if there is a particle occupying it, otherwise we call it \textit{free}.
Every cell can be occupied by at most one particle at all times.
We define the neighborhood of a particle $p$ as
$N(p) = ( \bigcup_{c \in C(p)} N(c) ) - C(p)$.
In Figure~\ref{fig:shapes}, the cells in the neighborhood of the particles are shown in gray.
Depending on the shape of a particle $p$ we have $|N(p)| \in \{6, 8\}$.
The position of a particle is uniquely defined by its \textit{head cell} $h \in \mathbb{Z}^2$.
The rotation of a particle is defined by a direction $r \in \mathbb{Z}_6$ that we call its \textit{orientation}.
In Figure~\ref{fig:shapes}, the orientation is depicted as an arrow inside the particle
and the head cell is the cell that contains this arrow.
The cells in the neighborhood of a particle are numbered
according to the orientation of the particle as shown in Figure~\ref{fig:shapes}.
The numbering starts with $0$ at the cell $n_r(h)$ and is increased in clock-wise direction around the particle.
We denote the neighboring cell of a particle with number $i$ as $n_i(p)$.
Finally, a \textit{configuration of a particle} is a tuple $(q, s, h, r)$
and consists of its state $q \in Q$, its shape $s \in S$,
its head cell $h \in \mathbb{Z}^2$, and its orientation $r \in \mathbb{Z}_6$.
Visually, any valid configuration of a particle can be achieved
by taking one of the configurations shown in Figure~\ref{fig:shapes},
rotating it by a multiple of $60^\circ$, and translating it to the desired position.
The \textit{configuration of a particle system} is the ordered set of configurations of all its particles.
We define two particles $u$ and $v$ to be \textit{connected} according to their configurations
if $N(u) \cap C(v) \neq \emptyset$ or, equivalently, $N(v) \cap C(u) \neq \emptyset$.
The configuration of a particle system $P$ then induces an undirected graph $G_P = (P, E)$
where $\{u, v\} \in E$ if and only if particles $u$ and $v$ are connected.
We call $G_P$ the \textit{connectivity graph} of the particle system $P$ according to its configuration.
A configuration is called \textit{connected} if the according connectivity graph is connected.

\begin{figure}
	\begin{minipage}{0.45\textwidth}
		\centering
		\includegraphics[scale = \figScale]{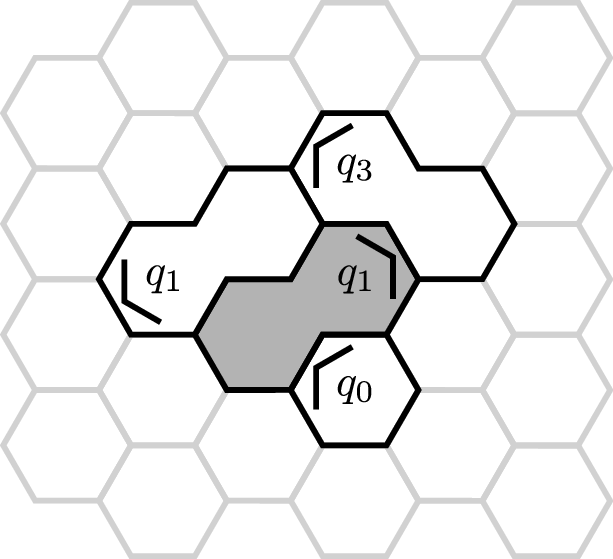}
	\end{minipage}%
	\hspace{0.1\textwidth}%
	\begin{minipage}{0.45\textwidth}
		\[(q_1, s_2, (n_0, \ldots, n_7))\]
		\[
			\begin{array}{lcl}
				n_0 = (q_3, s_2, \{6\})		& & n_1 = \epsilon 				\\
				n_2 = (q_0, s_1, \{0, 1\})	& & n_3 = \epsilon 				\\
				n_4 = \epsilon 				& & n_5 = (q_1, s_2, \{5, 6\})	\\
				n_6 = (q_1, s_2, \{5, 6\})	& & n_7 = (q_3, s_3, \{6\})
			\end{array}
		\]
	\end{minipage}
	\caption{
		Example of an element of the domain of the transition function.
		The transition function is to be evaluated for the gray particle in the left half of this figure.
		The labels in the head cells of the particles represent their respective state.
		The right half shows the formalization of the depicted situation.
	}
	\label{fig:config}
\end{figure}

\begin{figure}
	\centering
	\includegraphics[scale = \figScale]{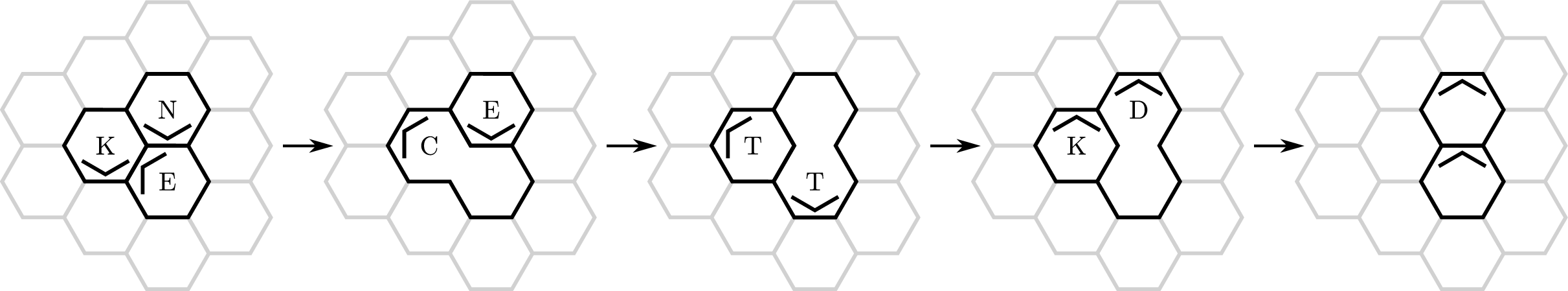}
	\caption{
		An example of a set of particles executing various actions over five rounds.
		The particles are labeled according to the actions they execute.
	}
	\label{fig:actions}
\end{figure}

Initially, the particle system is assumed to be in some connected configuration
in which all particles are of shape $s_1$ and in the start state $q_0$.
The configuration of the particle system has to stay connected at all times.
The particles act in synchronous rounds
according to their respective probabilistic transition function $\delta$ that has the following signature.
\[ \delta : Q \times S \times N^8 \rightarrow \mathcal{P}(Q \times A) \]
The transition function maps the state and shape of a particle $p$ combined with some neighborhood information
to tuples of a new state and an \textit{action}.
The neighborhood information represents the local view of $p$ on its neighboring cells $N(p)$ and
particles occupying these cells.
For a single cell it is of the following form.
\[N = (Q \times S \times \mathcal{P}(\{0, \ldots, 7\})) \cup \{\epsilon\}\]
Considering a specific cell $c = n_i(p)$, the $(i+1)$-th element of the tuple $N^8$ is $\epsilon$
if $i \geq |N(p)|$ or the cell $c$ is free.
Otherwise, i.e., if the cell is occupied by a particle $p'$,
it consists of the state and shape of $p'$ as well as the set $\{j \mid n_j(p') \in C(p)\}$.
An example of an element of the domain of the transition function is given in Figure~\ref{fig:config}.
The set of actions is defined as $A = \{N, T, E, C, D, K\}$
and the definitions of these actions are given in Table~\ref{tab:actions}.
\begin{table}
	\centering
	\input{tables/actions.tex}
	\caption{
		Definitions of the actions a particle of admissible shape can execute.
	}
	\label{tab:actions}
\end{table}
An example of particles executing these actions is depicted in Figure~\ref{fig:actions}.
Which actions a particle can execute depends on the shape of the particle;
the admissible shapes for each action are given in Table~\ref{tab:actions}.
If a transition function assigns an action to a particle in a shape that is inadmissible for the action,
the action fails and the transition function is not applied.
If a particle $p$ attempts to expand into a cell that is already occupied by a particle $p'$ in round $t$
and remains occupied by $p'$ in round $t + 1$, the expansion of $p$ fails and, again, the transition function is not applied.
If a particle $p$ attempts to expand into a cell that is occupied by a particle $p'$ in round $t$
but is freed in round $t + 1$ due to a contraction of particle $p'$, $p$ can expand into the cell.
In general, if multiple particles attempt to expand into the same cell,
one arbitrary particle succeeds in expanding while the others fail and their transition function is not applied.

\section{Research Challenges}
\label{sec:researchChallanges}
The proposed model allows to investigate various problems.
For example, the class of \textit{smart paint problems} might be considered.
Here, the surface of an object is to be covered as uniformly as possible by a set of particles.
A second example is the class of \textit{shape formation problems}
in which particles have to arrange to form specific shapes.
In \textit{bridging and covering problems} particles have to cover or bridge gaps in given structures.
For the above problem classes, particles are not allowed to execute the divide or kill action.
Another example is what we call the \textit{macrophage problem}.
This problem is inspired by biological cells called macrophages \cite{macrophages}
that can be found, for example, in the human body.
Their task is to engulf and digest pathogens.
To locate their target, macrophages use chemotaxis \cite{chemotaxis},
i.e., they move along a gradient of chemicals in their environment.
Our model can be extended to include chemotaxis, e.g., as in \cite{BEB08}.
The macrophage problem models the behavior of macrophages in the following way.
The macrophage and the pathogen are represented by two distinct particle systems;
accordingly the macrophage and the pathogen can move independently of each other.
The challenge now is to find an algorithm such that the macrophage hunts and surrounds the pathogen to immobilize it.
Several variants of this problem emerge
when deciding whether particles should be allowed to execute the divide and kill action
and whether macrophage and pathogen move at the same or different speeds.
For an additional variant, a new action could be added that allows the macrophage to kill particles
that belong to the pathogen.

Also, variants of our model might be of interest.
Firstly, in a physical realization particles may become faulty
and a particle system may not start in a well-initialized configuration.
For a particle system to handle such occurrences it may be necessary to allow
for particles to detect other faulty particles and to cope with them,
for example, by allowing particles to carry other particles.
Such a model would require self-stabilizing algorithms.
Secondly, the model may be extended from two to three dimensions using, for example,
the rhombic dodecahedron as the shape for the particles \cite{BCH00,YZLM01}.
Lastly, the model could be modified so that particles act asynchronously.

\newpage
\bibliographystyle{plain}
\bibliography{literature}

\end{document}

%% file: tables/actions.tex
\begin{tabular}{|c|c|c|p{9.5cm}|}
	\hline
	\textbf{Action} & \textbf{Description} & \textbf{Shape} & \textbf{Configuration Changes} \\ \hline \hline
	$N$ &
	Null Action &
	$\{s_1, s_2\}$ &
	None.
	\\ \hline
	
	$T$ &
	Turn &
	$\{s_1, s_2\}$ &
	If $s = s_1$ set $r \leftarrow r + 1$.

	Otherwise set $h \leftarrow n_{r + 3}(h)$ and then $r \leftarrow r + 3$.
	\\ \hline
	
	$E$ &
	Expand &
	$\{s_1\}$ &
	Set $h \leftarrow n_r(h)$ and $s \leftarrow s_2$.
	\\ \hline
	
	$C$ &
	Contract &
	$\{s_2\}$ &
	Set $s \leftarrow s_1$.
	\\ \hline
	
	$D$ &
	Divide &
	$\{s_2\}$ &
	Set $s \leftarrow s_1$ and then add a copy of the particle to the system
	with $h' \leftarrow n_{r + 3}(h)$ and $q'$ being the state after $\delta$ was applied.
	\\ \hline
	
	$K$ &
	Kill &
	$\{s_1\}$ &
	Remove the particle from the particle system.
	\\ \hline	
\end{tabular}

%% file: main.bbl
\begin{thebibliography}{10}

\bibitem{AAC+00}
H.~Abelson, D.~Allen, D.~Coore, C.~Hanson, G.~Homsy, T.~F. Knight, R.~Nagpal,
  E.~Rauch, G.~J. Sussman, and R.~Weiss.
\newblock Amorphous computing.
\newblock {\em Communications of the ACM}, 43(5):74--82, 2000.

\bibitem{Adl94}
L.~M. Adleman.
\newblock Molecular computation of solutions to combinatorial problems.
\newblock {\em Science}, 266(11):1021--1024, 1994.

\bibitem{AE07}
R.~Ananthakrishnan and A.~Ehrlicher.
\newblock The forces behind cell movement.
\newblock {\em International Journal of Biological Sciences}, 3(5):303--317,
  2007.

\bibitem{AAD+06}
D.~Angluin, J.~Aspnes, Z.~Diamadi, M.~J. Fischer, and R.~Peralta.
\newblock Computation in networks of passively mobile finite-state sensors.
\newblock {\em Distributed Computing}, 18(4):235--253, 2006.

\bibitem{BEB08}
L.~Bai, M.~Eyiyurekli, and D.~E. Breen.
\newblock An emergent system for self-aligning and self-organizing shape
  primitives.
\newblock In {\em Proceedings of SASO '08}, pages 445--454, oct. 2008.

\bibitem{BA99}
A.-L. Barab{\'a}si and R.~Albert.
\newblock Emergence of scaling in random networks.
\newblock {\em Science}, 286(5439):509--512, 1999.

\bibitem{BPEA+01}
Y.~Benenson, T.~Paz-Elizur, R.~Adar, E.~Keinan, Z.~Livneh, and E.~Shapiro.
\newblock Programmable and autonomous computing machine made of biomolecules.
\newblock {\em Nature}, 414(6862):430--434, 2001.

\bibitem{BCH00}
H.~Bojinov, A.~Casal, and T.~Hogg.
\newblock Emergent structures in modular self-reconfigurable robots.
\newblock In {\em Proceedings of ICRA '00}, volume~2, pages 1734--1741, 2000.

\bibitem{BDLS96}
D.~Boneh, C.~Dunworth, R.~J. Lipton, and J.~Sgall.
\newblock On the computational power of dna.
\newblock {\em Discrete Applied Mathematics}, 71:79--94, 1996.

\bibitem{BMV12}
V.~Bonifaci, K.~Mehlhorn, and G.~Varma.
\newblock Physarum can compute shortest paths.
\newblock In {\em Proceedings of SODA '12}, pages 233--240, 2012.

\bibitem{BKRT04}
Z.~J. Butler, K.~Kotay, D.~Rus, and K.~Tomita.
\newblock Generic decentralized control for lattice-based self-reconfigurable
  robots.
\newblock {\em International Journal of Robotics Research}, 23(9):919--937,
  2004.

\bibitem{CDBG11}
K.~C. Cheung, E.~D. Demaine, J.~R. Bachrach, and S.~Griffith.
\newblock Programmable assembly with universally foldable strings (moteins).
\newblock {\em IEEE Transactions on Robotics}, 27(4):718--729, 2011.

\bibitem{Chi94}
G.~Chirikjian.
\newblock Kinematics of a metamorphic robotic system.
\newblock In {\em Proceedings of ICRA '94}, volume~1, pages 449--455, 1994.

\bibitem{CP96}
G.~Chirikjian and A.~Pamecha.
\newblock Bounds for self-reconfiguration of metamorphic robots.
\newblock In {\em Proceedings of ICRA '96}, volume~2, pages 1452--1457, 1996.

\bibitem{DPSS11}
E.~D. Demaine, M.~J. Patitz, R.~T. Schweller, and S.~M. Summers.
\newblock Self-assembly of arbitrary shapes using rnase enzymes: Meeting the
  kolmogorov bound with small scale factor (extended abstract).
\newblock In {\em Proceedings of STACS '11}, pages 201--212, 2011.

\bibitem{Dol00}
S.~Dolev.
\newblock {\em Self-Stabilization}.
\newblock MIT Press, 2000.

\bibitem{EP07}
A.~Efrima and D.~Peleg.
\newblock Distributed models and algorithms for mobile robot systems.
\newblock In {\em Proceedings of SOFSEM '07}, pages 70--87, 2007.

\bibitem{FR03}
R.~Fitch and D.~Rus.
\newblock Self-reconfiguring robots in the usa.
\newblock {\em Japanese Robotic Society Journal}, 21(8):4--10, 2003.

\bibitem{FNKB88}
T.~Fukuda, S.~Nakagawa, Y.~Kawauchi, and M.~Buss.
\newblock Self organizing robots based on cell structures - cebot.
\newblock In {\em Proceedings of IROS '88}, pages 145--150, 1988.

\bibitem{Fuk02}
Y.~Fukui.
\newblock Mechanics of amoeboid locomotion: Signal to forces.
\newblock {\em Cell Biology International}, 26(11):933--944, 2002.

\bibitem{Ker12}
S.~Kernbach, editor.
\newblock {\em Handbook of Collective Robotics -- Fundamentals and Challanges}.
\newblock Pan Stanford Publishing, 2012.

\bibitem{Kle00}
J.~Kleinberg.
\newblock The small-world phenomenon: an algorithmic perspective.
\newblock In {\em Proceedings of STOC '00}, pages 163--170, 2000.

\bibitem{KSP12}
T.~Krasinski, S.~Sakowski, and T.~Poplawski.
\newblock Autonomous push-down automaton built on dna.
\newblock {\em Informatica}, 36:263--276, 2012.

\bibitem{LTT+10}
K.~Li, K.~Thomas, C.~Torres, L.~Rossi, and C.-C. Shen.
\newblock Slime mold inspired path formation protocol for wireless sensor
  networks.
\newblock In {\em Proceedings of ANTS '10}, pages 299--311, 2010.

\bibitem{McL08}
J.~McLurkin.
\newblock {\em Analysis and Implementation of Distributed Algorithms for
  Multi-Robot Systems}.
\newblock PhD thesis, Massachusetts Institute of Technology, 2008.

\bibitem{NKC03}
R.~Nagpal, A.~Kondacs, and C.~Chang.
\newblock Programming methodology for biologically-inspired self-assembling
  systems.
\newblock Technical report, AAAI Spring Symposium on Computational Synthesis,
  2003.

\bibitem{Ost04}
E.~Ostergaard.
\newblock {\em Distributed control of the ATRON selfreconfigurable robot}.
\newblock PhD thesis, University of Southern Denmark, 2004.

\bibitem{WWA00}
J.~E. Walter, J.~L. Welch, and N.~M. Amato.
\newblock Distributed reconfiguration of metamorphic robot chains.
\newblock In {\em Proceedings of PODC '00}, pages 171--180, 2000.

\bibitem{WWA04}
J.~E. Walter, J.~L. Welch, and N.~M. Amato.
\newblock Distributed reconfiguration of metamorphic robot chains.
\newblock {\em Distributed Computing}, 17(2):171--189, 2004.

\bibitem{WBX11}
Y.~Wang, P.~Brown, and Y.~Xia.
\newblock Nanomedicine: Swarming towards the target.
\newblock {\em Nature Materials}, 10(7):482--483, 2011.

\bibitem{WTTN11}
S.~Watanabe, A.~Tero, A.~Takamatsu, and T.~Nakagaki.
\newblock Traffic optimization in railroad networks using an algorithm
  mimicking an amoeba-like organism, physarum plasmodium.
\newblock {\em Biosystems}, 105(3):225--232, 2011.

\bibitem{WS98}
D.~J. Watts and S.~H. Strogatz.
\newblock Collective dynamics of 'small-world' networks.
\newblock {\em Nature}, 393(6684):440--442, 1998.

\bibitem{cellSignaling}
Wikipdedia.
\newblock Cell signaling.
\newblock \url{http://en.wikipedia.org/wiki/Intercellular_communication}, 2013.

\bibitem{chemotaxis}
Wikipdedia.
\newblock Chemotaxis.
\newblock \url{http://en.wikipedia.org/wiki/Chemotaxis}, 2013.

\bibitem{complexNetworks}
Wikipdedia.
\newblock Complex networks.
\newblock \url{http://en.wikipedia.org/wiki/Complex_network}, 2013.

\bibitem{macrophages}
Wikipdedia.
\newblock Macrophages.
\newblock \url{http://en.wikipedia.org/wiki/Macrophages}, 2013.

\bibitem{WLWS98}
E.~Winfree, F.~Liu, L.~A. Wenzler, and N.~C. Seeman.
\newblock Design and self-assembly of two-dimensional dna crystals.
\newblock {\em Nature}, 394(6693):539--544, 1998.

\bibitem{YSS+07}
M.~Yim, W.-M. Shen, B.~Salemi, D.~Rus, M.~Moll, H.~Lipson, E.~Klavins, and
  G.~S. Chirikjian.
\newblock Modular self-reconfigurable robot systems.
\newblock {\em IEEE Robotics Automation Magazine}, 14(1):43--52, 2007.

\bibitem{YZLM01}
M.~Yim, Y.~Zhang, J.~Lamping, and E.~Mao.
\newblock Distributed control for 3d metamorphosis.
\newblock {\em Autonomous Robots}, 10(1):41--56, 2001.

\end{thebibliography}
